# On the Unfathomableness of Consciousness by Consciousness

*Why do physicists widely agree on the assured extent of their professional knowledge, but not so philosophers?*
*An exchange of letters and a discussion with Carl Friedrich von Weizsäcker*

Klaus Gottstein

On the occasion of Carl Friedrich von Weizsäcker's 81$^{st}$ birthday, a colloquium was held on July 3, 1993 in Niederpöcking on the Starnberg Lake. One of the topics was "The epistemological foundation of physics from Kant to von Weizsäcker". Walter Schindler and Thomas Görnitz gave the introductory speeches. I took part in the discussion which followed. Afterwards I got the impression that I had not succeeded in making my points clear. Therefore, on July 5, I wrote an explanatory letter to Prof. von Weizsäcker. This was the beginning of a correspondence on the questions intimated by the title above. In addition, a private discussion between Prof. von Weizsäcker and myself took place. Some readers may be interested in the answers which Carl Friedrich von Weizsäcker has given to the naïve questions and views of a former physicist with hardly any education in philosophy. The letters and my notes on our discussion follow in chronological order. (I am, of course, responsible for any errors in my representation of von Weizsäcker's remarks in my notes.)

July 5, 1993

Dear Mr. von Weizsäcker,

I am afraid that my improvised remarks in the discussion after the speeches of Messrs. Schindler and Görnitz were not sufficiently clear. My first comment concerned the remarkable circumstance that the physicists of a given historical period agree to a large extent on the observable phenomena of nature and on their underlying laws, while in philosophy, there are always different schools with different conceptions of truth and which accuse each other of error. To me, this appears to be due to the fact that physics is concerned with phenomena which present themselves in the same way to all human beings possessing the same brains and the same state of consciousness as conditioned by the historical era under consideration. In philosophy, however, the brain seeks to discover which possibilities of cognition exist for the brain. Isn't that so as if a man wants to take measure of himself? This can never lead to unambiguous results, I should think! One cannot help remembering Münchhausen who tore himself out of the swamp by his own pigtail. There is, perhaps, a connection between this difficulty of the human brain in grasping the conditions of its capacity for gaining knowledge and in representing these conditions unambiguously and free of contradiction, and Gödel's Theorem on the impossibility of proving the consistency of a formal system by its own means (see, for example: Alfred Gierer: Physics, Life and Mind; Piper Verlag 1985). It seems that the realm of religions, that means of faith, begins at this point.

My second remark concerned the often cited connection between Kant's inclusion of the subject into objective reality in the course of its recognition, and quantum theory where one cannot abstract from the observer (Mr. Schindler spoke about this). It appears to me that this is not a real connection, but rather a kind of parallelism. After all, in quantum theory it is only shown that the experimenter with his apparatus, in observing atomic processes, interferes with the process to be observed (by emitting light quanta and by other means), so that the process cannot be observed objectively. But this is of no consequences under macroscopic conditions. Also according to quantum mechanics, we can still speak of macroscopic objects observed independently of the subject, the experimenter; in astronomy, for example. Kant, however, had in mind the macroscopic world when setting out his epistemology, which seeks to determine what can be known of reality to the human subject. He knew nothing yet about atomic processes. Therefore, quantum theory and the <u>Critique of Pure Reason</u> deal with different matters, which nevertheless have in common that the observing subject becomes entangled with the observed phenomena – in the case of Kant in a very general way as the human subject which is dependent on its sensory experiences and for which the 'thing-in-itself' is inaccessible anyway and, therefore, meaningless; in quantum theory, however, the observing subject is entangled only for observations of processes in domains where classical physics is no longer valid.

Would you agree?

   Cordially,

   Klaus Gottstein

                          July 12, 1993

Dear Mr. Gottstein,

I thank you very much for having more fully explained the sense of your contribution to the discussion. I must confess, however, that I see the problem in a slightly different way. It is a particularity of life that living beings are capable of performing many activities, without being able to explain what they are really performing, or even without reflecting on what they are performing. With animals this is plainly so, but it is also largely so with humans. Now, science is one of the great cultural discoveries in which one has seen that man is capable of very complex activities when certain methods are applied; for example, empirical and mathematical methods. Anyone who then reflects more closely on what was really done here encounters the difficulty that he acted like animals do who are able to do something without being able to provide the information why that is so.

This is already implied in your parable of the brain which wants to understand the brain itself. The question asked here, on the other hand, is an entirely meaningful question, and particularly so when it does not attempt to solve all great problems at once, but toils on the problems of understanding which arise from science anyway and which prove to be important. Thomas Kuhn has differentiated between normal science as puzzle-solving under a given paradigm, that is, under a given example of solving riddles, and scientific revolutions in which one is forced to change paradigms. Scientific revolutions are mostly made possible through a reflection on what one has really done, a reflection which one may very well call philosophical. In this sense, for example, the philosophical reflection of Ernst Mach on the meaning of classical mechanics contributed significantly to Einstein's being able to discover the theory of relativity and Heisenberg's being able to discover quantum mechanics. Only when normal science resumes fully its activity under a new paradigm, it is no longer necessary to tackle the more difficult question which I have called the philosophical one.

Moreover, I believe that the connection between Kant and quantum theory can play an explanatory role after all though, as you correctly say, it is not identical at all with these problems. Kant's question was how experience was actually possible. Thereby he clarified the important role of the judgement of the scientific subject in understanding scientific objects. Precisely this question came up again in quantum theory in the context of special examples. That is, methodically Kant asks similar questions, the concrete contents, on the other hand, are different. I might add that, in the context of quantum theory, classical physics has been cast into a new light as a sort of limiting case. Kant's problem which, of course, referred only to classical physics, is indeed cast into a new light by quantum theory. But I shall not try to detail these matters in this letter.

Unfortunately, I am a little short of time and I cannot provide a more complete account. I have written on this issue in some of my books; perhaps one day we shall be able to find an opportunity for a quiet discussion.

       Cordially,

       CF Weizsäcker

October 5, 1993

Dear Mr. von Weizsäcker,
......................
It is certainly correct that humans and other living beings are capable of many things without knowing how they have done it. But this cannot mean that something can be done which is logically impossible. When the brain wants to understand the brain - and not that in a medical-physiological sense in which the human brain is an object as other objects of natural science, but in the philosophical sense as an organ that experiences the world – then it is true that this desire is legitimate. But, to choose another analogy, isn't this like the eye desiring to look at itself without the help of a mirror? So it is not really surprising that in philosophy there are as many answers to one question as there are philosophical schools.

Accidentally, I found a description of Karl Jasper's philosophy in the book by Simone Hersh: "The Philosophical Marvel".

*"... Objectivity, therefore, is much more than just a scientific requirement: it is of an ontological nature. Just because of this, Man strives to grasp the world objectively as a totality. But he cannot escape the fundamental situation of all thinking: that he, as the thinking subject, faces an object-reality which he wants to raise to a totality. And here he is astonished: just this is impossible. He can never be included in this presumed totality insofar as he thinks himself as a subject. The world will never combine for him into a totality. ... The thinker, therefore, always remains in the Kantian subject-object split and in a world which can never become for him a unity and a totality. We cannot reach the knowledge of totality. It is true that we can always proceed toward a horizon which all the time surmounts what we already know; but we can never proclaim a knowledge claiming that it refers to the totality. Totality is denied to us by our situation as recognizing subjects. ..."*

This quotation seems to express in a much more perfect way what I have tried to say with my simple examples of the yardstick which attempts to measure itself, the brain which wants to understand itself, and the eye, which strives to see itself without a mirror.

Cordially

Klaus Gottstein

November 3, 1993

Dear Mr. Gottstein,

...........

The topic which you raise again in your letter is basically too difficult for a brief discussion by an exchange of letters. We should talk quietly about it, the two of us, which I will be glad to do.

For the time being let me say that even Jaspers' very careful formulation does not convince me at all. Perhaps it is consistent in the tradition of a thinker greater than Jaspers, namely Kant. On the other hand, I know that there have been thinkers in the history of human thought of the same greatness as Kant who look at these matters very differently. From antiquity I would name Plato; from Asia both the Vedanta philosophy and Buddha. All of these do not see in knowledge the contraposition of a subject with precise knowledge of itself with a recognized object in such a way that, as Jaspers means, the subject can never become the object of cognition. Definitely not. This distinction, in all its rigor, is only due to Descartes, if I understand correctly, and I would say that already Fichte, Schelling and Hegel after Kant sensed that a transgression is needed here, and a different view. But, as I said, this is a wide field and we should discuss these matters orally.

Cordially,

Carl Friedrich Weizsäcker

December 6, 1993

Dear Mr. von Weizsäcker,

At the outset, with regard to your letter of November 3, perhaps already this remark: The fact that great thinkers have seen the possibilities of human self-knowledge in very different ways was exactly the starting point of my remarks at the colloquium discussion on the occasion of your last birthday. If the human brain would be able to analyze its own essence in an irrefutable manner, all serious philosophers would have to agree on the possibilities of human self-knowledge, just as all serious physicists basically agree on the theory of relativity and on quantum mechanics. The fact that such an agreement does not exist among the great philosophers, appears to me to confirm that man is overburdened with the task of self-analysis in a similar way as the yardstick with the task of self-measurement. It would require an external yardstick. Without such an external yardstick, "arbitrary" results are obtainable. The yardstick external to man, however, would not belong to the realm of science any more, but to that of religion. After all, the thinkers whom you have named in your letter, Plato, Buddha and the Vedanta philosophers, may perhaps be attributed to this realm..

With the help of the "religious yardstick" man can recognize himself and his individual relation to divinity, and this in a dimension which is no longer directly accessible to reason, to science. Religious revelation has taken different forms in different cultural areas. This non-unambigiousness religion and philosophy have in common. It would be an interesting question what this commonality means. For non-religious persons an answer readily at hand is that religion is just as imagined by man as is philosophy. But what answer should believers give? Again, a ready-made answer is: all religions except one's own are imagined by men; only one's own religion is founded by God. For the adherents of the major world religions this is, no doubt, the "correct" answer in each case. The question remains what consequences we have to draw from this. In one of your books and during one of our discussions you said that you would have become, of course, a Buddhist or a Muslim if you had not happened to be born in Europe, but had been born in East Asia or in the Middle East. For this reason, and because of the wisdom of the great Asian religions, you could not adhere to the claim of Christianity to be the only salvation-bringing religion, as Karl Barth has done. This answer has never satisfied me from a logical point of view, since you and I were not born and raised in East Asia or in the Middle East but in Europe in the same spiritual world as Karl Barth. Therefore, to us the commandments of our religion should be valid, including the "Go ye therefore, and teach all nations and baptize them …". This does not exclude that, seen from a higher level, the representatives of other religions have received the order to convert *us*. If this were so, it would not be our concern. For us, there would remain only one explanation: God's ways are not our ways. We have a different message.

Cordially,

Klaus Gottstein

March 9, 1994

Dear Mr. von Weizsäcker,

Since our last discussion and my letter of December 6, 1993, I have continued to search for philosophical statements regarding the possibility of man to know himself – not only as an object, but in the totality of one's own self. The search is not yet finished, the literature is too extensive. But I want to give you an intermediate report as this may save time during our upcoming conversation.

I do not intend to return to Karl Jaspers who thought that the perceiver cannot attain the knowledge of totality, as there is always only a horizon which discloses itself. In your letter of November 3 you have termed Jasper's statement as Kantian but not convincing. I have come across the following statements, which I can give here as a kind of anthology, for which I beg your pardon:

Bertrand Russell: " The value of philosophy consists, on the contrary, essentially in its uncertainty, which is inherent in her."
*(quoted from Wilhelm Weischedel, 34 große Philosophen in Alltag und Denken).*

Karl Popper and John Eccles: "We have to face the fact that we live in a world in which almost everything which is truly important remains essentially unexplained"
*(quoted from Volker Spierling, Kleine Geschichte der Philosophie).*

Maurice Merleau-Ponty considers as hopeless the endeavor to penetrate to a theoretically fixed structural entity of basic human experiences. … Philosophy must be inherent in the original, indissoluble union ("mixture") of man and the world which no analysis can circumvent. Starting from a finiteness of man, newly understood and positively grasped in this way, philosophy consummates itself as hermeneutics of experience or interpretation of existence, respectively. It renounces the concept that the world, things and consciousness could be generally determined and become adequately fixed by thinking. *(Margot Fleischer in: Philosophen des 20. Jahrhunderts, Wissenschaftliche Buchgesellschaft, 3. Auflage 1992).*

Merleau-Ponty speaks of the "paradox of time": *"... I cannot really take possession of my time before I understand myself completely, and this moment can never come, because, if it came, it would still be just a moment surrounded by a horizon of the future, which itself would stand in need of unfolding in order to be understood. ... The reflection ... clarifies everything except its own role." (quoted according to Erich Christian Schröder, in: Margot Fleischer (Hrsg.), Philosophen des 20. Jahrhunderts).*

Michel Foucault emphasizes the impossibility to reach behind our entanglement with the world through a reflection on the subject, let alone with the intention of providing an ultimate substantiation. No longer is a reason common to all held as the governor for knowledge and truth. *(quoted from: Erich Christian Schröder, in: Margot Fleischer (Hrsg.), Philosophen des 20. Jahrhunderts).*

According to Michel Foucault, human activity, life and language develop their own, quasi-transcendental structure of laws which man struggles in vain to dominate. The attempt to think the unthought and to catch up with the continuously escaping origin leads to an endless duplication in empiricism and transcendentalism which does not help us advance at all. *(Bernhard Waldenfels in: Margot Fleischer (Hrsg.), Philosophen des 20. Jahrhunderts).*

For Theodor Adorno, the patient acceptance of the heterogeneous, dispersed and refractory given is essential to philosophy. Philosophy interprets it, seeks for traces of hope in it, tries to unravel its ciphers, but without entertaining the illusion that world could be discoverable or moldable into a unity of meaning, or could open a view to a "world behind".
(Margot Fleischer, loc. cit.).

The thesis of W.V.O. Quine is: There exist total theories of the world which deeply differ from each other, yet are empirically equivalent in the sense that they imply exactly the same observation-categorical principles. *(Felix Mühlhölzer in: Margot Fleischer, loc. cit.).*

With Ernst Bloch, the starting point is "being", that existence from time immemorial which we always are already and which we can never get hold of because as the existing one it remains unfathomably ahead of thinking though itself pushing and driving us. … Bloch has circumscribed the non-constructibility of the absolute question which we are to ourselves, in the well-known formula: "darkness of the presently lived moment."  We live, I am, but just this directness of life, of the "being" which carries us and from which everything wells up, cannot be caught up with, neither experiencing nor understanding. *(Wolfdietrich Schmied-Kowarzik, in: Margot Fleischer, loc. cit.).*

This is as far as I have got. But perhaps these samples are already sufficient to show that the limits of self-knowledge are felt also by modern, not at all religious, thinkers. The philosopher Jeanne Hersch, a disciple of Jaspers, I must admit, has expressed this in the following words: "Existence knows that it has not created itself" (Jeanne Hersch, The Philosophical Marvel). This is very close to statements such as the one by Gregor of Nazianz (330-390), one of the Early Fathers of the Church: God in his being is inaccessible for human thought. *(quoted from Wolf-Dieter Hauschild, Gregor von Nazianz, in: Klassiker der Theologie, Beck-Verlag)*

….
        Cordially,

        Klaus Gottstein

K. Gottstein

Notes on the conversation with Carl Friedrich von Weizsäcker on March 16, 1994

Just to finish this subject and then proceed to questions more important to me I recalled the second topic of my letter of July 5, 1993 (the parallel between Kant's inclusion of the subject into the objective reality in the process of perceiving it, and the statements of quantum mechanics) and mentioned that Kant's statement, of course, referred to the macroscopic world, whereas quantum mechanics leads to differences from classical physics only in the microscopic world, so that there is no real connection between Kant and quantum mechanics. CFvW, however, stayed with this topic for a while, it is especially close to him. He mentioned the dispute between Boltzmann and Mach on the existence of atoms, a dispute based on the fact that even the behavior of macroscopic bodies (energy, temperature, the second law of thermodynamics) cannot be understood without contradictions if one supposes on the basis of classical physics that macroscopic bodies are composed of atoms. Since Mach realized this, he denied Boltzmann the right to assert the existence of atoms. In this context I stated that the necessity to apply quantum mechanics here to macroscopic bodies is due to the inclusion of the concept of microscopic atoms into the consideration. – CFvW agreed with this and added that the chemists of the 19$^{th}$ century had such great success with the notion of atoms because, in their happy naiveté, they did not see the associated complications. Moreover, CFvW maintained that the study of Kant could definitely contribute to the understanding of quantum mechanics. He, CFvW, had just written a paper on Kant and quantum mechanics. – Regarding the relation of science and philosophy, CFvW pointed to Heidegger whom he had known well and who had remarked that scientists do not think. They don't reflect on why their methods are successful, what they do with these methods and what the consequences are of their procedures. It is not self evident, said CFvW, that the application of mathematics describes nature "correctly". But scientists prefer not to think about this, because they can obtain a university chair more easily by producing scientific results then by writing poor philosophical treatises. I objected that one could just as well write good philosophical treatises and then obtain a chair in philosophy. CFvW replied to my remark, which obviously referred to his own career, that this was very difficult.

The discussion then turned to the first topic of my letter of March 9, 1994: the question whether total self-knowledge is possible. Jaspers and the philosophers I had cited in my letter had denied this possibility. CFvW indicated that he was not at all convinced or impressed by these citations. One had to know, he said, that these are philosophers of the 20$^{th}$ century. Jaspers shows philosophical arrogance when he says that he knows that the ego cannot know itself. Other centuries came to totally different conclusions. He pointed especially to Descartes who had shown that everything can be doubted except that one doubts, and therefore exists. From this point Descartes started in constructing his philosophy. He even derived a proof for the existence of God which, however, is the weakest point in his system of thought. He, CFvW, said that he did not share Descartes' philosophy but he could defend Descartes as an attorney, so to speak. He added that he developed his own philosophy from a dispute with Descartes. In any case, Descartes showed that one cannot doubt one's own consciousness.

I raised two objections:

1. Jaspers and the other philosophers which I had cited did not doubt their own existence but the possibility of grasping the totality of their own selves.

2. The centuries named by CFvW, in which philosophers could supposedly understand their own selves, are only the few centuries of the Age of Enlightenment. In the many centuries before, philosophers were of a different opinion. I mentioned Thomas Aquinas and the Early Fathers of the Church who understood man as a creature of God and God as inaccessible to reason. Faith in revelation had priority over philosophical wisdom which does not contribute anything to salvation.

CFvW replied that with the Gentile Christians (i.e. Christians of non-Jewish origin, of which Thomas Aquinas was one), two different cultures encountered each other: the Greek thinking of Aristotle, and Christianity which arose from Jewish sources. In order to reconcile both, Thomas made the compromise to give precedence to Christian revelation. At his time, CFvW said, Thomas could not do more. For having a unified view of the world without compromises he had been born either 700 years too late or 900 years too early. Even today religion and enlightenment are still incomplete. The next steps must lead both sides to the realization that each of them needs the other one for its own perfection (religion the enlightenment as well as enlightenment the religion).

In this context CFvW mentioned that Niels Bohr with his philosophy of complementarity is probably the most important thinker of the $20^{th}$ century. He is the Socrates of our time who, like Socrates, proclaims that he knows nothing in order to show to others who pretend to know that they, too, know nothing. Then, when a solution is asked for, the reply is: "We shall talk more tomorrow". Bohr's philosophy of complementarity allows to view an object under different – even seemingly incompatible – aspects. Here is an indication of how the philosophy of the future could look.

I then mentioned that the present epoch, in any case, lacks a world view determining the ethical behavior of men. Although many wise books have been written – among them those by CFvW - none of these are evaluated in the sense that they would become sufficiently effective in politics. Christianity once played the role of the dominant world view in Europe, today it is no longer binding for the majority of people. Marxism which had been the dominant world interpretation for several decades has collapsed. The peaceful coexistence of men requires a common ethical basis. I asked if one could "invent" an ethics because it is needed for practical purposes, as Hans Küng has tried to do. Isn't ethics a result of a firmly rooted world interpretation? Can one generate ethics without this foundation?

CFvW objected that Küng sets out from the statement: "No world peace without peace between religions!" – I asked if this does not represent an overestimation of religion. Are religions still taken so seriously that a conclusion of peace between the leaders of different religions can still affect world politics? But I admitted that religious differences can provide the pretext for expulsions, civil wars and other belligerent actions. But wouldn't politicians find other reasons when religious pretexts were no longer available? (Remark at the writing of these notes: For a long time now there has been peace between the Catholic and Protestant leaderships, but which effect does this have in Northern Ireland?)

CFvW defended Küng's "Project World Ethos." It did not represent an ethics on command. After all, Küng worked out the common ethical foundations of the world religions by detailed investigations of which the results are laid down in several books. In this research he found, i.a., that the "Golden Rule" is valid everywhere: What you do not want anybody to do to yourself, do not do to anybody! In addition, there are common elements in mysticism and meditation.

At this point, CFvW inserted a biographical note: At the age of twenty, on the occasion of a Protestant church service, he noticed the preacher's great pathos. He then realized that this pathos served to drown his own disbelief: The preacher did not believe what he said. At a Catholic service, the Latin mass and the solemn singing had given him, CFvW, the impression: This is the tradition of two thousand years, this I can bear. He had thought of converting to the Catholic faith. But there was the dogma of infallibility of the Pope when he spoke ex cathedra. Had he already been a Catholic, then he would not have left because of this. But to submit to this dogma voluntarily through conversion would have been too great a sacrifice.

Here I, too, made a biographical remark: I was molded by the Protestant youth circle in the parish of pastor Martin Niemöller in Dahlem at the time of the struggle of the Church with the Nazi state. I have met preachers who believed what they said. Today, their belief would perhaps be termed fundamentalist, but it enabled believers to risk their lives in offering resistance.

CFvW remarked that he admired these believers and their actions. But he had not taken part in this. Niemöller and Gollwitzer he had known very well. He had met Niemöller in 1958 at the synod in Spandau in which he had participated as an invited expert. The topic was atomic armament. Opinion was divided among the members of the synod, the Church was tried to the breaking point. Niemöller proved to be the preserver of unity. Although he himself was a radical pacifist, he had learned as church president to respect the opinions of differently minded persons. He called on the quarreling parties to reassemble under the gospel in spite of their divergences. This appeal had its effect.

After this excursion, we came back to Hans Küng and the problem of cooperation among the religions in the elaboration of a common ethics. CFvW said that he supports Küng and only regrets that Küng is not on good terms with the Pope. Catholics, of course, should be included in this cooperation. To him as a Protestant school boy sitting in class with Catholics and Jews, it had already seemed strange that the contingency of birth decides on one's faith. If his parents had been Catholics or Buddhists, he would have become a Catholic or a Buddhist. It could not be true that God had given the right faith only to one religion and had left all others in error.

I replied that with such reflections he, CFvW, placed himself in the role of God. The Christian has a clear mandate: "Go ye therefore, and teach all nations and baptize them!" (compare the letter of December 6, 1993)

CFvW responded that lack of tolerance is the main shortcoming of Christians. A Buddhist could be Christian simultaneously without any problem. Why is the inverse impossible? CFvW's Japanese Christian acquaintances (in Japan a small minority) had told him that it is not their aim to convert all Japanese to Christianity; rather it is their aim to make the Japanese people Buddhists again. To him the wording of the dogmas is not essential, CFvW said.

I objected that then he could not really speak the Christian Confession of Faith. CFvW replied that in this matter he followed his great grandfather who had been a liberal professor of protestant theology. Someone had told his great grandfather that the Confession of Faith must be changed since one knows now that several statements therein are not correct. His great grandfather had replied that he was against a modification of the Confession of Faith but that he would propose to allow anyone when speaking the Confession to think what he considers to be correct. The Bible contains several contradictions, CFvW said. He had discovered the following contradiction in the Old Testament: At one point (Second Book of Samuel, Chapter 24), David's census is said to be God's order, and the observance of this order is subsequently punished by God. This seems to be the original version. The later chronicler (First Book of Chronicles, Chapter 21) has apparently modified the text in order to erase this contradiction. Now Satan inspires King David with the idea of the census, and God punishes.

For the history of creation, CFvW said, he had constructed for himself the following anecdote which he thought had a high degree of veracity. To the Rabbi who around 500 B.C. in Babylon writes down the history of creation comes a young man and asks him if he really believes that God has created the world in six days. The Rabbi answers: "Don't you see that this story is a parable?" The texts of the Old Testament which partly originate around 900 B.C., were apparently, according to recent findings, edited and compiled in an "editorial conference" around 500 B.C.

CFvW returned again to Kant. Logically, the existence of God could not be deduced from Kant's system. But Kant presupposed the existence of God for moral reasons and restricted his system so that there remained room for the belief in God.

CFvW quoted what many of his Buddhist acquaintances had told him: Europeans do not realize what damage they cause with their Aristotelian logic: they ruin the world.

I replied that people like the Babylonians, Assyrians and Huns who had known nothing about Aristotle, had also inflicted terrible damages. To this CFvW answered that the Europeans with their Aristotelian method of thinking had finally defeated these people. Moreover, the reproach of Buddhists against Europeans is not that of cruelty; rather it concerns the thoughtlessness of their actions with respect to the consequences.

To this I said that grim consequences are in most cases collateral effects or long term consequences of measures with good intentions. In order to be able to prevent such undesirable collateral effects and long term consequences, the possibility of their occurrence must be forecast, and for this interdisciplinary and often international cooperation is necessary. What must we do today? I stressed that, regardless of the unclarified philosophical and theological questions, what now matters is to avoid catastrophes. I mentioned the early efforts by CFvW within the the Federation of German Scientists and the Starnberg Institute, and the Federation of American Scientists and the Pughwash Movement. CFvW said that the atomic bomb was the alarm signal, indicating to us that we cannot proceed in the same way. The message has been heard, the Third World War has so far been avoided.

I mentioned once again my hope that a social theory and a world view for the modern era could be found – perhaps in the form of CFvW's anticipated mutual approach of religion and science.

March 22, 1994

Dear Mr. von Weizsäcker,

In the meantime I have read with great interest your essay : "On the Structure of Physics: Kant and Quantum Theory", which you mentioned to me during our conversation on March 16. To my surprise, you have not only discussed there the question of whether there is a genuine connection between Kant's thoughts on the role of the subject in the observation of "objective reality" and the concepts resulting from quantum mechanics, or whether this is only a parallelism. But on the last page of the essay you make also statements regarding the second topic of our discussion, namely the question whether it is possible for consciousness to grasp itself in its totality. In our conversation, you designated as philosophy of the 20th century the collection of statements by modern philosophers from my letter of March 9 (expressing the view that man cannot reach ultimate clarity on himself) which should be confronted with the quite different findings of earlier centuries. In particular, you mentioned Descartes who had inferred his existence from the indubitable fact of his doubt. I had remarked in this context that also the cited philosophers of the 20th century did not dispute the certitude of their own existence. After all, the knowledge of one's own being cannot be equated to grasping the totality of one's own consciousness.

Now I see on page 8 of your essay on Kant and quantum theory that, in speaking of Descartes, you characterize a certitude of self-consciousness which goes beyond the absolute certitude of self-existence, with the words "that may be" which express a lower level of certainty. At the end of the relevant paragraph, you write: "I find myself in the thinking and observing consciousness and interpret this in the frame of evolution." After all, when one finds oneself somewhere, then one does not know how one got there and has to rely on interpretations.

In the last paragraph of page 8 of your essay you write: "The problem (of mind and form) cannot be solved in a rational-speculative way. It presupposes, according to my conviction, religious experience: ethical practice and meditative perception." Do I interpret correctly that – much in the sense of the statements in my letters of July 5, October 5 and December 6 of 1993 – you come to the conclusion that there are insurmountable barriers for the rational knowledge of the brain by the brain, therefore of the totality of the self by the self, and that behind these barriers lies the domain of religion, in which the last questions can be answered by faith?
….
       Cordially,

       Klaus Gottstein

June 29, 1994

Dear Mr. Gottstein
…………
Your letter of March 22 is still on my desk, unanswered. To blame for this is the fact that I receive vast amounts of mail these days. …

The pecularity is, as it appears to me, that both of us, you and I, have basically a very similar conception of human consciousness in the context of nature, but that again each of us spontaneously chooses ways of expression in which the other cannot join and which provokes the other to contradiction.

From my point of view, for example, my main stumbling-stone with respect to your manner of expression has been when you say that there are limits to the rational cognition of the brain by the brain. First, it should be said that in the tradition, as it exists since Descartes, nobody could have spoken like that, because for Descartes consciousness, the res cogitans, was of a totally different substance than matter, the res extensa, to which the brain belongs. In my understanding of Descartes, he held the opinion that consciousness fundamentally knows itself. The question how this relates to the brain was difficult for him. According to his conviction, animals which also have a brain after all, were automatic machines without any consciousness. On the other hand, he could not deny the link between matter and consciousness for humans. He placed it into a causal mutual influencing in the only unpaired organ in the human head, the pineal gland. For Descartes it would have been self-evident that the brain cannot recognize itself, because the brain is just matter and not thinking substance.

Now, this is not my personal opinion at all. I merely wanted to say that the history of the philosophy of consciousness up to Kant is not understandable if this division by Descartes is not taken into consideration and that the mind-body problem, which still confronts biologists today, is a consequence of this Cartesian separation.

When one attempts to remove this separation, when one simply denies its existence, and this would be my interpretation especially also of quantum theory, then indeed it becomes almost self-evident that consciousness cannot know itself completely; I like to express this by the formula: "consciousness is an unconscious act". This, however, I mean at first again in a descriptive way, and for this I do not necessarily need the recourse to the brain. It is sufficient that one knows, for example, psychoanalysis since Freud. Fundamentally, also earlier thinkers, for example Goethe, have known this quite well. When I then tried to describe the role of the brain in this context, then I first had to start from the fact that according to quantum theory, at least as I interpret it, already the atom or the elementary particle is part of something virtually conscious. 'Virtually' means that it may take several billions of years before a consciousness arises which can even reflect on itself. Schelling said nature is the mind which does not recognize itself as mind. In this way of speaking I may perhaps say of man that the brain is not the carrier of consciousness, but the brain is an aspect of consciousness which one gets to see when one approaches it from the outside, that is spatially.

Whether or not this solves our controversy is not clear to me, but I wanted to make at least these few remarks.

       Cordially

       Carl Friedrich Weizsäcker

                                             August 1, 1994

Dear Mr. von Weizsäcker,

I thank you very much for your letter of June 29, 1994 which I found in my mail after longer journeys. I am very glad to take it from your lines that, strictly speaking, we agree in substance and that you have only disapproved of my manner of expression, which did not sufficiently take into account the tradition since Descartes, although you, too, dissociate yourself from Descartes' separation of thinking and material substance. Your statement that, after removing this separation, it becomes almost self-evident that consciousness cannot know itself completely, and my statement that the brain cannot know itself completely for logical reasons (in my letter of July 5, 1993 I had suspected a link with Gödel's theorem of incompleteness) show to me satisfactory agreement. This is so because, actually, I meant human consciousness when I spoke of the "brain". Strictly speaking, I only wanted to say that *man* cannot understand himself completely. You seem to agree with this. When I spoke of the brain, this was a concession to natural science which apparently has identified the brain as the seat of consciousness. When the brain is destroyed, there is no more consciousness, even if the heart continues to beat for a long time. But recourse to the brain is not important for me either.

As you will remember, the starting point of our discussion was my remark that the conspicuous inability of philosophers to agree on the deepest elements of human existence including the basic questions of ethics – which is tantamount to the occurrence of different schools of philosophy feuding with one another – could perhaps be traced back to the fundamental inability of man to attain self-knowledge. Physicists, on the other hand, can reach consent, in an equally remarkable way, on the results of their measurements and on whether or not these results agree with the ideas of theory. They just have to deal with res extensa. Here the yardstick does not intend to measure itself, but objects which are external to it, and this is possible.

There remains only the question what consequences can be drawn from this realization. Is philosophy, in as much as it gives statements on the totality of human existence, only an expression of the prevailing social circumstances and views, interesting enough as such, but not a gateway to truth? I am inclined to give an affirmative reply to this question.
….

       Cordially,

       Klaus Gottstein